\documentclass[prl,superscriptaddress,twocolumn]{revtex4}

\usepackage{enumerate}
\usepackage{amsfonts,amssymb,amsmath}
\usepackage[]{graphics,graphicx,epsfig}
\usepackage{amsthm}

\usepackage{graphicx}
\usepackage{dcolumn}
\usepackage{natbib}
\usepackage{color}
\usepackage{multirow}
\usepackage{ulem}

\begin{document}


\title{On Quantum Semipermeable Barriers: Investigating Maxwell's Demon Toolbox}

\author{Andrzej Grudka}
\affiliation{Institute of Spintronics and Quantum Information, Faculty of Physics, Adam Mickiewicz University, 61-614 Pozna\'n, Poland}

\author{Pawe{\l} Kurzy{\'n}ski}
\email{pawel.kurzynski@amu.edu.pl}
\affiliation{Institute of Spintronics and Quantum Information, Faculty of Physics, Adam Mickiewicz University, 61-614 Pozna\'n, Poland}
\affiliation{Centre for Quantum Technologies, National University of Singapore, 117543 Singapore}

\author{Antoni W{\'o}jcik}
\affiliation{Institute of Spintronics and Quantum Information, Faculty of Physics, Adam Mickiewicz University, 61-614 Pozna\'n, Poland}

\date{\today}


\begin{abstract}
We study quantum Maxwell's demon in a discrete space-time setup. We consider a collection of particles hopping on a one-dimensional chain and a semipermeable barrier that allows the particles to hop in only one direction. Our main result is a formulation of a local unitary dynamics describing the action of this barrier. Such dynamics utilises an auxiliary system $\mathcal{A}$ and we study how properties of $\mathcal{A}$ influence the behaviour of particles. An immediate consequence of unitarity is the fact that particles cannot be trapped on one side of the barrier forever, unless $\mathcal{A}$ is infinite. In addition, coherent superpositions and quantum correlations are affected once particles enter the confinement region. Finally, we show that initial superposition of $\mathcal{A}$ allows the barrier to act as a beam splitter.

\end{abstract}

\maketitle


\section{Introduction} 

Maxwell's demon stimulates research in fundamental physics for more than 150 years. Among many, let us mention its fruitful contributions to the physics of information \cite{Landauer,Bennett,MDR} and quantum mechanics \cite{QMD1,QMD2,QMD3,QMD4,QMD5}. The demon's scenario is usually formulated in terms of particles moving inside a partitioned box and an observer, the demon itself, who operates the slit in the partition to allow the transfer from the left part to the right one, but to prevent the transfer in the opposite direction. The workings of the demon are still an object of vivid academic discussions since they constitute a simple example capable of grasping the essence of many fundamental issues like observation, control and irreversibility. 

In this work we revisit the quantum Maxwell's demon scenario. The action of the demon is equivalent to the workings of a semipermeable barrier and the dynamics implemented by such a barrier is inevitably irreversible. However, any irreversible dynamics can be described as an extended dynamics on the primary system $\mathcal{S}$ (particles in the box) and an auxiliary one $\mathcal{A}$ (demon's memory). The goal of this paper is to analyse the properties of $\mathcal{A}$ and the consequences of its interactions with $\mathcal{S}$. To simplify the analysis we use a discrete space-time model, since in our case the problem of discreteness and continuity is not a physically relevant issue. Moreover, such an approach allows for an algorithmic treatment of dynamics in terms of sequential operations applied to the system, hence it is along the lines of quantum information-processing.

The motivation to study the unitary evolution of $\mathcal{S}+\mathcal{A}$ comes from at least two reasons. It is commonly believed that the fundamental laws of nature are reversible and the observed irreversible processes stem from not tracing all elements of a bigger system. It is therefore crucial to understand how many additional elements one needs to trace in order to observe reversible {\it inner workings} behind an apparent irreversibility. In addition, by understanding the inner workings of the Maxwell's demon we make a step towards construction of artificial demons, microscopic systems capable of transforming information into work \cite{Ueda}.

Our model allows us to make the following observations. First, we find that particles cannot be confined in a finite location for an infinite amount of time, unless $\mathcal{A}$ is capable of storing an infinite amount of information. If $\mathcal{A}$ is finite, the particles can be trapped for a limited period of time. Next, we study how coherent superpositions and quantum correlations within $\mathcal{S}$ are affected by interaction with $\mathcal{A}$. Finally, we show that an initial superposition of $\mathcal{A}$ provides a resource for the barrier to act as a beam splitter and that this resource is consumed once a particle's state becomes a spatial superposition.


\section{Preliminaries}


\subsection{Basic model}

\begin{figure}
	\includegraphics[trim=80 620 80 0,clip,width=8cm]{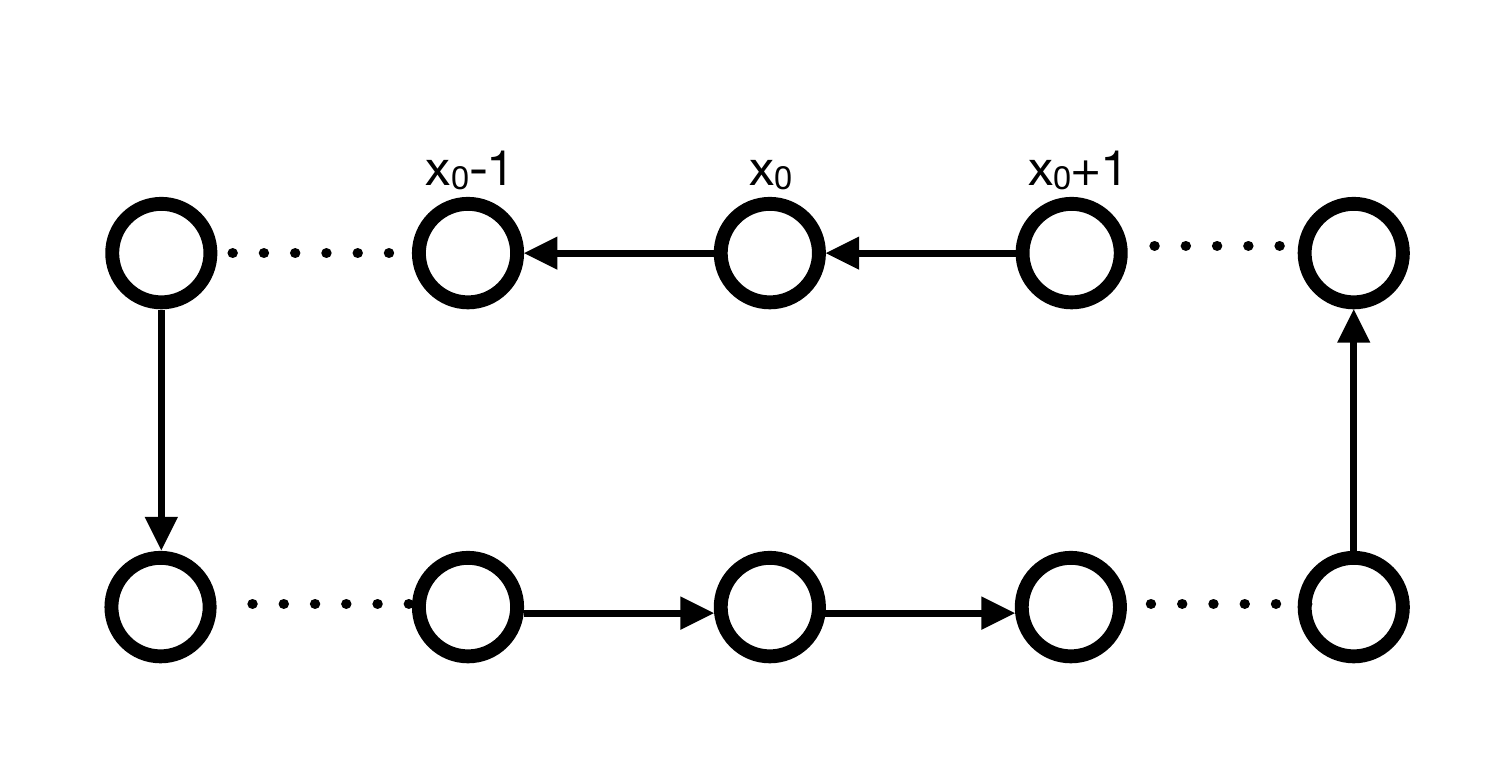}
  \caption{Schematic representation of the free motion generated by the operation $U$ (\ref{U}).} \label{f1}
\end{figure}

Due to the discrete space-time formulation our system can be modelled by a discrete-time quantum walk (DTQW). DTQW's are quantum counterparts of classical random walks and classical lattice gas automata \cite{Aharonov,Meyer}. Here we focus on a simple lattice -- a one-dimensional chain. The state of a single particle is described by two variables, the position $x\in \{1,2,\ldots,M\}$ and the direction of movement $c\in \{\leftarrow,\rightarrow \}$ (we follow the DTQW convention and call it a {\it coin}). The positions $x=1$ and $x=M$ correspond to the left and the right boundaries of the box, respectively. At time $t$ the system is in a state
\begin{equation}\label{state}
|\psi_t\rangle = \sum_{x=1}^M \left(\alpha_{x,t}|x,\rightarrow\rangle + \beta_{x,t}|x,\leftarrow\rangle \right),
\end{equation} 
where the probability amplitudes obey the normalization condition $\sum_x (|\alpha_{x,t}|^2 + |\beta_{x,t}|^2)=1$. The system evolves in steps
\begin{equation}
|\psi_{t+1}\rangle = U|\psi_{t}\rangle,
\end{equation}
where the unitary evolution operator is a conditional translation 
\begin{eqnarray}
& &U |x, \rightarrow\rangle = |x+1, \rightarrow\rangle, ~~(x<M) \nonumber \\
& &U |x, \leftarrow\rangle = |x-1, \rightarrow\rangle, ~~(x>1) \nonumber \\
& &U |1, \leftarrow\rangle = |1, \rightarrow\rangle, \nonumber \\ 
& &U |M, \rightarrow\rangle = |M, \leftarrow\rangle. \label{U}
\end{eqnarray}
The above is just a free motion inside the box -- the particle constantly jumps in one direction determined by the state of the coin (see Fig. \ref{f1}) and bounces off the barriers, in which case the state of the coin flips. Note, that such dynamics is classical in a sense that it transforms the basis states into the basis states. Any superposition results solely from an initial state preparation. 

In case of $N$ non-interacting particles the system's state is described by
\begin{equation}
|\psi_t^{(N)}\rangle = \sum_{x_1,c_1,\ldots,x_N,c_N}\alpha_{x_1,c_1,\ldots,x_N,c_N,t}|x_1,c_1\rangle\otimes\ldots\otimes |x_N,c_N\rangle.
\end{equation}
Although the particles can be in principle distinguishable, for the demon it is irrelevant which particle is which. Moreover, we assume that initially no two particles occupy the same state and that the dynamics will never evolve two, or more, particles into the same state. Therefore, we can describe the system with a help of fermionic creation operators $a_{x_c}^{\dagger}$. The $N$-partite state at time $t$ becomes
\begin{equation}\label{indist1}
|\psi_t^{(N)}\rangle=\sum_{x_1,c_1,\ldots,x_N,c_N}\alpha_{x_1,c_1,\ldots,x_N,c_N,t}a_{{x_1}_{c_1}}^{\dagger}\ldots a_{{x_N}_{c_N}}^{\dagger}|\mathbf{0}\rangle,
\end{equation} 
where $|\mathbf{0}\rangle$ is the vacuum state and the coefficients $\alpha_{x_1,c_1,\ldots,x_N,c_N,t}$ are antisymmetric. Its evolution is given by,
\begin{equation}\label{multiU1}
|\psi_{t+1}^{(N)}\rangle = U_N|\psi_t^{(N)}\rangle,
\end{equation}
where $U_N$ transforms each creation operator according to the following rule 
\begin{eqnarray}
& &a_{x_{\rightarrow}}^{\dagger} \rightarrow a_{x+1_{\rightarrow}}^{\dagger},~~(x<M) \nonumber \\ 
& &a_{x_{\leftarrow}}^{\dagger} \rightarrow a_{x-1_{\leftarrow}}^{\dagger}, ~~(x>1) \nonumber \\
& &a_{1_{\leftarrow}}^{\dagger} \rightarrow a_{1_{\rightarrow}}^{\dagger}, \nonumber \\
& &a_{M_{\rightarrow}}^{\dagger} \rightarrow a_{M_{\leftarrow}}^{\dagger}. \label{multiU2}
\end{eqnarray}


\subsection{Statement of the problem}

In simple words, the basic problem investigated by us is the following: a particle evolves according to the above model and we want to find a way to trap it in a region $\mathcal{R} = \{x|1 \leq x \leq x_0\}$. This problem seems trivial, since an intuitive solution is to wait for the particle to enter $\mathcal{R}$ and then to close the entry at $x_0$ by inserting an impenetrable barrier that reflects particles on both sides (flip their coin states if the particle is at $x_0$). Indeed, this would be trivial if one knew the initial state of the particle in advance. The initial state would determine the time interval, $t\in [t_{in},t_{out}]$, during which the particle passes through $\mathcal{R}$. It would be enough to keep the entries open for $t<t_{in}$ and closed for $t>t_{in}$.  

The problem gets more complicated if the initial state is unknown. In this case one has to employ an agent, the demon, to operate the entry. The demon determines the particle's state and then it picks the right time to insert the barrier. Moreover, in case there is more than one particle, the demon needs to let the new particles in and to prevent the old ones from getting out. From now on we will refer to the demon as the semipermeable barrier. The goal is to find a unitary evolution that describes its action.

Semipermeable barriers contract the effective dimension of the system. They transform many different states into the same one (Appendix A) and their action can be described by Kraus operators (Appendix B). However, since our goal is to find out its unitary description, we focus on the properties of an auxiliary system $\mathcal{A}$ that, together with the original system $\mathcal{S}$, undergoes a unitary evolution. What is important, throughout the work we assume that external observers do not have access to $\mathcal{A}$. The unitary workings of the barrier need to be reflected in the properties of $\mathcal{A}$, in its own dynamics and in its interaction with $\mathcal{S}$. 

In particular, we look for the answers to the following questions:
\begin{enumerate}
\item What unitary dynamics on $\mathcal{S}+\mathcal{A}$ traps a particle in a finite region $\mathcal{R}$? 
\item How the dynamics of particles depend on the size of $\mathcal{A}$?
\item What happens to superpositions and quantum correlations within $\mathcal{S}$ once particles enter $\mathcal{R}$?
\item How initial superposition of $\mathcal{A}$ affects the dynamics of $\mathcal{S}$?
\end{enumerate}


\section{Results} 


\subsection{Reversibility}

\begin{figure}
	\includegraphics[trim=80 620 80 0,clip,width=8cm]{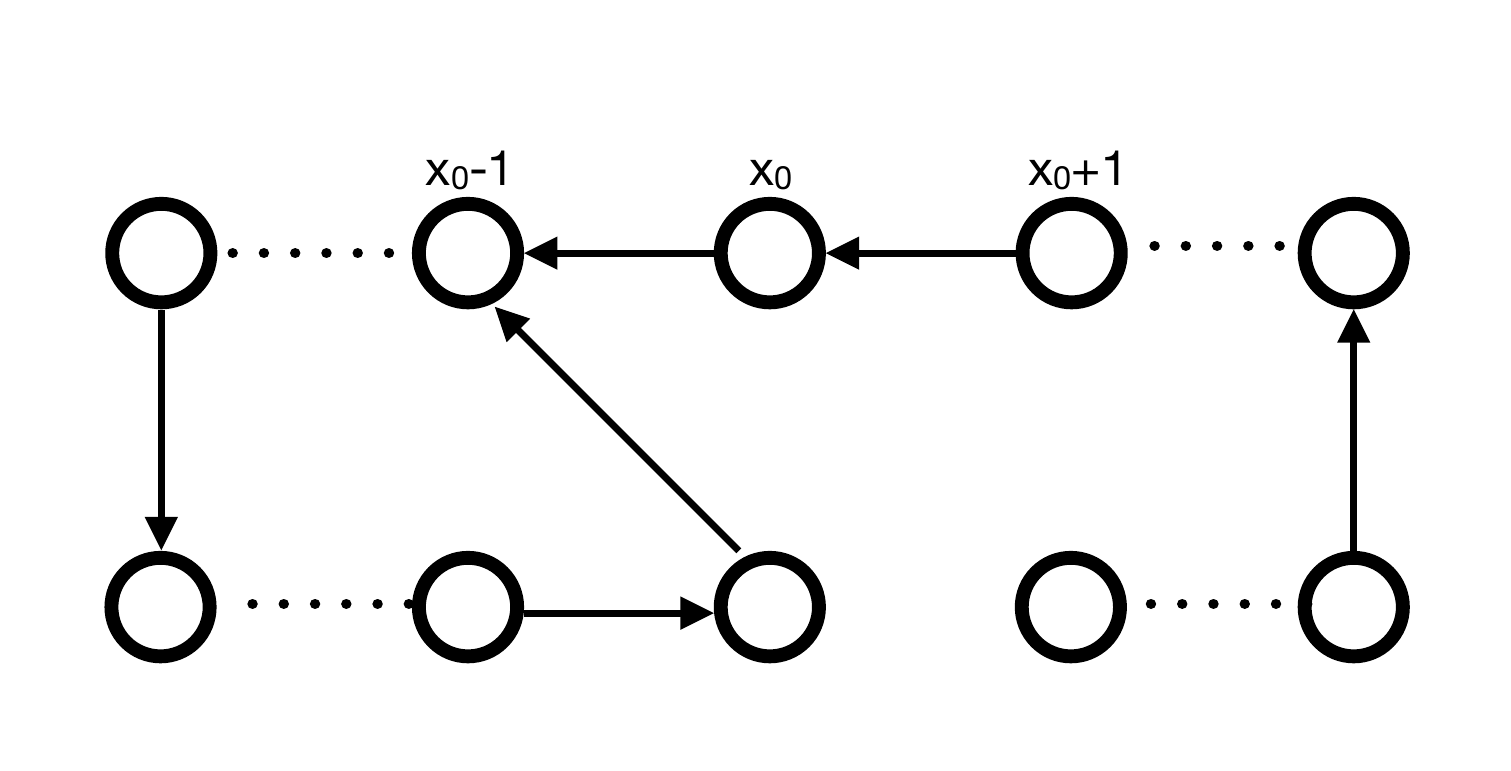}
  \caption{Schematic representation of the dynamics with a semipermeable barrier located at $x_0$. The barrier allows the particle to pass in only one direction.} \label{f2}
\end{figure}

We are looking for a unitary operator $V$ on a joint system $\mathcal{S}+\mathcal{A}$ capable of trapping particles in a finite region $\mathcal{R}$. We assume that $V$ is local, i.e., its nontrivial action is limited to $x_0$. The whole dynamics is described by
\begin{equation}\label{totevol}
|\Psi_{t+1}^{(\mathcal{S}+\mathcal{A})}\rangle = (U\otimes \openone_{\mathcal{A}})V|\Psi_t^{(\mathcal{S}+\mathcal{A})}\rangle,
\end{equation}
where $U$ is given by (\ref{U}), $\openone_{\mathcal{A}}$ is the identity operator acting on $\mathcal{A}$, and  $|\Psi_t^{(\mathcal{S}+\mathcal{A})}\rangle$ is the joint state of $\mathcal{S}$ and $\mathcal{A}$. A schematic representation of the desired transformation is presented in Fig. \ref{f2}

To determine $V$, let us first focus on a single particle and observe that the above dynamics is deterministic and reversible. This means that we can analyse the system evolving both, forward and backward in time. If we evolve it forward in time, an external observer sees a particle moving constantly in one direction till it reaches $\mathcal{R}$. From this moment the particle moves there and back within the boundaries of $\mathcal{R}$. There is no randomness -- just a deterministic contraction of the state space. If we evolve the system backward in time, an external observer sees a particle bouncing off the $\mathcal{R}$'s boundaries till at some seemingly undetermined moment it breaks free and leaves $\mathcal{R}$. This looks like a random behaviour.

The time-reversibility and determinism imply that the system has to have a memory of the particle's past. This memory must be stored on $\mathcal{A}$. More precisely, the particle's entry time into $\mathcal{R}$ is encoded on $\mathcal{A}$. Therefore, if the evolution were reversed, $\mathcal {A}$ would determine the particle's exit time. This means that $\mathcal{A}$ acts as a meter that counts the time the particle spends inside $\mathcal{R}$. 


\subsection{Where is $\mathcal{A}$?}

The next problem is the physical location of $\mathcal{A}$. In principle it can be a particle's internal degree of freedom. The semipermeable barrier could use this degree of freedom to flag each particle that enters $\mathcal{R}$. However, its state would have to change each time the particle bounces off the semipermeable barrier. Otherwise the evolution would be irreversible (see Appendix A). The change of $\mathcal{A}$'s state would count the time the particle spends inside $\mathcal{R}$. 

The above choice would imply a peculiar effect. Consider two identical particles. One of them enters $\mathcal{R}$ at time $t_1$ and the other one at time $t_2$. If the difference $t_2-t_1 \geq 2x_0$ then, once inside $\mathcal{R}$, the two particles would have different internal states. They would become effectively distinguishable. Any test, such as Hong-Ou-Mandel effect \cite{HOM}, would fail to confirm their indistinguishability. 

In principle, we could accept effective distinguishability of particles inside $\mathcal{R}$, however note that the action of the semipermeable barrier described by Kraus operators does not lead to such an effect (Appendix C). We prefer to construct a model which recovers the Kraus operator formalism once $\mathcal{A}$ is traced out. That is why we abandon the possibility that $\mathcal{A}$ is associated with particles. This implies that $\mathcal{A}$ must be an external system interacting with particles. The interaction must occur at $x_0$, since we insist that the action of the barrier is local.


\subsection{What is $\mathcal{A}$ and how it works?}

Let us still focus on only one particle. The goal of $\mathcal{A}$ is to make the particle change the direction of its movement each time it is about to leave $\mathcal{R}$. The following transformation on $\mathcal{S}$ must be implemented
\begin{equation}
|x_0,\rightarrow\rangle \rightarrow |x_0,\leftarrow\rangle.
\end{equation} 
The particle must also have a possibility to enter $\mathcal{R}$, which implies another transformation
\begin{equation}
|x_0,\leftarrow\rangle \rightarrow |x_0,\leftarrow\rangle.
\end{equation} 
These two transformations change two different states into the same one. To do this in a reversible way, the semipermeable barrier needs at least an auxiliary two-level system, a qubit, whose basis states are labelled $|0\rangle$ and $|1\rangle$. With a help of an auxiliary qubit one can find a unitary operation $W$ whose action is
\begin{eqnarray}
W |x_0,\leftarrow\rangle \otimes |0\rangle &=& |x_0,\leftarrow\rangle\otimes |0\rangle, \nonumber \\
W |x_0,\rightarrow\rangle\otimes|0\rangle &=& |x_0,\leftarrow\rangle\otimes|1\rangle. \label{W}
\end{eqnarray}
In general, due to the locality assumption, the operation $W$ acts effectively in a four-dimensional space spanned by the states $|x_0,\rightarrow\rangle\otimes|0\rangle$, $|x_0,\rightarrow\rangle\otimes|1\rangle$, $|x_0,\leftarrow\rangle\otimes|0\rangle$ and $|x_0,\leftarrow\rangle\otimes|1\rangle$. All the remaining states are unchanged by its action, i.e,
\begin{equation}
W|x,c\rangle\otimes|q\rangle = |x,c\rangle\otimes|q\rangle~~~\text{if}~~~x\neq x_0,
\end{equation} 
where $c=\leftarrow,\rightarrow$ and $q=0,1$. Moreover, $W$ is only a part of the operation $V$ (see (\ref{totevol})), that describes the action of the barrier. The full form of $V$ will be given in a moment.

Now, let us check if one qubit is enough. Once the above transformation is implemented, the particle travels inside $\mathcal{R}$ and the qubit remains in the state $|1\rangle$. After $2x_0-1$ steps the system evolves into $|x_0,\rightarrow\rangle\otimes|1\rangle$. However, the unitarity of $W$ implies that 
\begin{equation}\label{ucon}
W|x_0,\rightarrow\rangle\otimes|1\rangle = \alpha |x_0,\rightarrow\rangle\otimes|0\rangle + \beta |x_0,\rightarrow\rangle\otimes|1\rangle,
\end{equation}
therefore the particle exits $\mathcal{R}$ after $2x_0 - 1$ steps. We conclude that, to achieve its goal, the semipermeable barrier has to have access to many qubits in the state $|0\rangle$. 

In Eq. (\ref{ucon}) $\alpha$ and $\beta$ are arbitrary. We fix
\begin{eqnarray}
W |x_0,\leftarrow\rangle \otimes |1\rangle &=& |x_0,\rightarrow\rangle\otimes |0\rangle, \nonumber \\
W |x_0,\rightarrow\rangle\otimes|1\rangle &=& |x_0,\rightarrow\rangle\otimes|1\rangle,\label{wallreversed}
\end{eqnarray}
but any other choice is good. Our choice implies the following symmetry -- if the initial qubit's state is changed from $|0\rangle$ to $|1\rangle$, the action of the semipermeable barrier is reversed, i.e., it allows to pass a particle incoming from the left and reflects a particle incoming from the right. 


\subsection{Size of $\mathcal{A}$ and finite trapping time}

The particle bounces off the semipermeable barrier every $2x_0-1$ steps and each time it does so a new qubit in the state $|0\rangle$ is needed. The reason why the bounce happens after $k(2x_0-1)$ steps, not $k2x_0$, is that the single step of the evolution (\ref{totevol}) consists of two consecutive operations, $V$ and $(U\otimes\openone_{\mathcal{A}})$. The first one transforms $|x_0,\rightarrow\rangle$ into $|x_0,\leftarrow\rangle$ and the second one transforms it further into $|x_0-1,\leftarrow\rangle$. Therefore, the trapped particle can never be observed in the state $|x_0,\leftarrow\rangle$ (see Fig. \ref{f2}). 

The number of time steps the particle spends inside $\mathcal{R}$ is proportional to the number of qubits available to the barrier. More  precisely, $k$ qubits will allow to trap the particle for $(k+1)(2x_0 -1)$ steps (we will show this in a moment). In principle, if the size $M$ of the box was known, it would be enough to use the barrier for $2(M-x_0)$ steps. After that time a standard barrier could be placed at $x_0$. However, the goal is to construct a universal semipermeable barrier capable of trapping the particle for any $M$, hence $M$ could be infinite.  In addition, putting a standard barrier would require an external intervention, in which case the dynamics would be time-dependent and the total system would no longer be autonomous. We therefore use a $k$-qubit string and accept the fact that particles are trapped for a finite period of time. Note, that in reality only a finite number of qubits is available.


\subsection{Unitary trapping}

Let $\mathcal{A}$ be a string of $k$ qubits whose initial state is 
\begin{equation}\label{zeros}
\underbrace{|0\rangle \otimes |0\rangle \otimes \ldots \otimes |0\rangle}_{k}.
\end{equation} 
We assume that the first qubit in the string is the one that is used by $W$ to implement the transformation (\ref{W}). Next, let $T$ be a $k$-qubit unitary operation that cyclically permutes the qubits, i.e., 
\begin{equation}\label{T}
T\left(|\phi_1\rangle\otimes \ldots |\phi_{k-1}\rangle \otimes |\phi_k\rangle \right) = |\phi_2\rangle \otimes \ldots |\phi_k\rangle \otimes |\phi_1\rangle,
\end{equation}
where $|\phi_i\rangle$ is the state of the i-th qubit. It is clear that $T^k = \openone^{\otimes k} = \openone_{\mathcal{A}}$. $T$ acts on $\mathcal{A}$ if and only if the particle is in the state $|x_0,\leftarrow\rangle$. The goal of $T$ is to replace the first qubit by the second one whenever the first one is used by $W$, i.e., whenever the particle bounces off the semipermeable barrier, or enters $\mathcal{R}$. In addition, the used qubit is moved to the last position and the previously used qubits, that are already at the end of the string, are shifted one position to the left. As a result, the information about the particle's past is stored at the end of the string. 

The total evolution of $\mathcal{S}+\mathcal{A}$ (\ref{totevol}) can be written as
\begin{eqnarray}\label{totdyn2}
|\Psi_{t+1}^{(\mathcal{S}+\mathcal{A})}\rangle = (U\otimes \openone^{\otimes k})V|\Psi_{t}^{(\mathcal{S}+\mathcal{A})}\rangle,
\end{eqnarray}
where 
\begin{eqnarray}
V &=& (\Pi_{T}\otimes T)(W\otimes\openone^{\otimes k-1}) \label{V} \\
&+& ((\openone_{\mathcal{S}} - \Pi_{T})\otimes \openone^{\otimes k})(W\otimes\openone^{\otimes k-1}), \nonumber
\end{eqnarray}
and $\Pi_{T} = |x_0,\leftarrow\rangle\langle x_0,\leftarrow|$. Of course, the above works for only one particle in the system. The multipartite case will be considered in a moment.

Let us follow the dynamics generated by (\ref{totdyn2}). Upon the particle's entry to $\mathcal{R}$ the first qubit is used and moved to the last position of the string. The transformation (\ref{W}) implies that its state remains $|0\rangle$. After the next $2x_0-1$ steps the next qubit is used, but this time the particle bounces off the barrier and the qubit's state is changed to $|1\rangle$. In general, after $(j-1)(2x_0-1)$ steps inside $\mathcal{R}$ the qubit string is in the state
\begin{equation}\label{string}
\underbrace{|0\rangle\otimes\ldots\otimes|0\rangle}_{k-j ~ unused} \otimes \underbrace{|0\rangle\otimes|1\rangle\otimes\ldots\otimes|1\rangle}_{j~used}.
\end{equation}
Note, that the first $|0\rangle$ state in the {\it used} section marks the particle's entry time. If the evolution was reversed, this $|0\rangle$ state would make the particle leave $\mathcal{R}$ after $(j-1)(2x_0-1)$ reversed steps.  

All qubits in the string are used once the particle spends $(k-1)(2x_0-1)$ steps in $\mathcal{R}$. At that moment the string is in the state 
\begin{equation}
|0\rangle\otimes|1\rangle\otimes\ldots\otimes|1\rangle.
\end{equation}
The particle makes one more round and after $k(2x_0-1)$ steps it bounces off the barrier for the last time. By doing so the particle reuses the first qubit and changes it's state to $|1\rangle$, so the string becomes 
\begin{equation}\label{ones}
|1\rangle\otimes|1\rangle\otimes\ldots\otimes|1\rangle.
\end{equation}
Now, due to (\ref{ucon}), the particle has to leave $\mathcal{R}$ upon the next arrival at the barrier. Therefore, the particle spends $(k+1)(2x_0-1)$ steps inside $\mathcal{R}$, during which the state of the qubit string is transformed from (\ref{zeros}) to (\ref{ones}).

Before we proceed, let us discuss one issue. In general, the qubit string can take one of $2^k$ different orthogonal states. In our model we use only $k+1$ of them. Therefore, in principle $\mathcal{A}$ could be compressed to a $(k+1)$-level system, or a much more efficient trapping procedure could be used to allow the particle to stay in $\mathcal{R}$ for $2^k(2x_0-1)$ steps. Still, remember that the above discussion concerns only a single-particle case. In a moment we will show that the whole set of $2^k$ states will be used in the multipartite scenario.


\subsection{Possible paradox and its resolution}

After $k(2x_0-1)$ steps inside $\mathcal{R}$ the qubit string is in a state $|1\rangle\otimes\ldots\otimes |1\rangle$ and one may wonder what would happen if we flipped it back to $|0\rangle\otimes\ldots\otimes |0\rangle$. In principle we can do the flipping every $k(2x_0-1)$ steps. Does it mean that the particle can be trapped in $\mathcal{R}$ forever?

The answer is negative due to the following unitarity property, already mentioned in this work. Any unitary transformation conserves the dimension of the state space. Note, that the number of different orthogonal states of the $\mathcal{S}+\mathcal{A}$ system inside $\mathcal{R}$ is fixed and finite. In particular, the effective state space of $\mathcal{S}+\mathcal{A}$ within $\mathcal{R}$ is $(k+1)(2x_0-1)$-dimensional. On the other hand, the number of states outside of $\mathcal{R}$ depends on the size of the box $M$, which can be arbitrarily large. If the particle was trapped for $T>(k+1)(2x_0-1)$ steps, the trapping operation would have to be contractive. This is because $T$ states (from outside of $\mathcal{R}$) would have to be transformed into at most $(k+1)(2x_0-1)$ states (inside $\mathcal{R}$). This would violate the unitarity.     


\subsection{Many particles}

Next, we generalize the previous transformation to the multipartite case. This time we are going to use fermionic creation operators, as in (\ref{indist1}). Let us first modify the $W$ operation. As before, due to the locality assumption, its action is non-trivial at $x=x_0$. This time this position can be occupied by zero, one, or two particles (one in the state $|x_0,\rightarrow\rangle$ and the other one in $|x_0,\leftarrow\rangle$). We assumed that particles are fermionic, therefore when two of them are at $x_0$, one that attempts to enter $\mathcal{R}$ and one that attempts to stay in $\mathcal{R}$, only one of them will end up in $\mathcal{R}$ in the state $|x_0-1,\leftarrow\rangle$. The other one must end up outside of $\mathcal{R}$ in the state $|x_0+1,\rightarrow\rangle$. This corresponds to the original Maxwell's demon situation in which two particles are heading for the slit, one from the inside and one from the outside. The demon is puzzled because it can decide either to keep the slit closed and reflect both particles, or to keep the slit opened and allow both particles to go through. Both possibilities produce equivalent effects (due to the indistinguishability of particles). In our case, we choose the second possibility. Of course, the problem can be generalized to include bosonic particles, in which case both can be trapped, but we do not consider it here.  

Let us consider $N$ particles in the system. The basis states of $\mathcal{S}+\mathcal{A}$ are
\begin{equation}
(a^{\dagger}_{{x_1}_{c_1}} \ldots a^{\dagger}_{{x_N}_{c_N}} )|\mathbf{0}\rangle\otimes|q_1 q_2 \ldots q_k\rangle,
\end{equation}
where $a^{\dagger}_{{x_i}_{c_i}}$ creates a particle in a state $|x_i,c_i\rangle$ ($1 \leq x_i \leq M$ and $c_i = \leftarrow,\rightarrow$) and $q_j = 0,1$. In the above we assume $x_i \leq x_{i+1}$. We also used the shorthand notation $|q_1\rangle\otimes\ldots\otimes|q_k\rangle \equiv |q_1 \ldots q_k\rangle$. The corresponding $N$-partite version of $W$ causes the following transformations of the basis states
\begin{eqnarray}
W_N ( \ldots a^{\dagger}_{{x_0}_{\leftarrow}}  \ldots )|\mathbf{0}\rangle\otimes|0\ldots\rangle = ( \ldots  a^{\dagger}_{{x_0}_{\leftarrow}}  \ldots )|\mathbf{0}\rangle\otimes|0  \ldots \rangle,  \nonumber \\
W_N ( \ldots a^{\dagger}_{{x_0}_{\rightarrow}}  \ldots )|\mathbf{0}\rangle\otimes|0 \ldots \rangle = ( \ldots a^{\dagger}_{{x_0}_{\leftarrow}} \ldots )|\mathbf{0}\rangle\otimes|1\ldots \rangle, \nonumber \\
W_N ( \ldots  a^{\dagger}_{{x_0}_{\leftarrow}}  \ldots )|\mathbf{0}\rangle\otimes|1 \ldots \rangle = ( \ldots  a^{\dagger}_{{x_0}_{\rightarrow}}  \ldots )|\mathbf{0}\rangle\otimes|0 \ldots \rangle, \nonumber \\
W_N ( \ldots a^{\dagger}_{{x_0}_{\rightarrow}}  \ldots )|\mathbf{0}\rangle\otimes|1 \ldots \rangle = ( \ldots a^{\dagger}_{{x_0}_{\rightarrow}} \ldots )|\mathbf{0}\rangle\otimes|1  \ldots \rangle, \nonumber \\ \label{mpW1}
\end{eqnarray}
if only one particle is at $x=x_0$. If two particles are at $x=x_0$, the action of $W_N$ is
\begin{eqnarray}
& & W_N ( \ldots a^{\dagger}_{{x_0}_{\leftarrow}} a^{\dagger}_{{x_0}_{\rightarrow}} \ldots )|\mathbf{0}\rangle\otimes|q_1 \ldots \rangle = \nonumber \\
& &~~~~~ ( \ldots a^{\dagger}_{{x_0}_{\leftarrow}} a^{\dagger}_{{x_0}_{\rightarrow}} \ldots )|\mathbf{0}\rangle\otimes|q_1  \ldots \rangle. \label{mpW2}
\end{eqnarray}
Finally, if there are no particles at $x=x_0$, the action of $W_N$ is 
\begin{eqnarray}
& &W_N(a^{\dagger}_{{x_1}_{c_1}} \ldots a^{\dagger}_{{x_N}_{c_N}} )|\mathbf{0}\rangle\otimes|q_1 \ldots \rangle = \nonumber \\ 
& &~~~~~(a^{\dagger}_{{x_1}_{c_1}} \ldots a^{\dagger}_{{x_N}_{c_N}} )|\mathbf{0}\rangle\otimes|q_1 \ldots \rangle.   \label{mpW3}
\end{eqnarray}
In the last two situations $W_N$ does not change the state of $\mathcal{S}+\mathcal{A}$.

The cyclic permutation of the qubit string can be done with the help of the previously introduced transformation $T$ (\ref{T}). This allows us to represent the action of the semipermeable barrier, capable of trapping more than one particle, as
\begin{eqnarray}
V_N &=& \left(a^{\dagger}_{{x_0}_{\leftarrow}} a_{{x_0}_{\leftarrow}}\otimes T \right)W_N \nonumber \\
&+& \left(a_{{x_0}_{\leftarrow}} a^{\dagger}_{{x_0}_{\leftarrow}}\otimes \openone_{\mathcal{A}} \right)W_N, \label{VN}
\end{eqnarray}
hence the one step of the total $N$-partite evolution is
\begin{equation}
|\Psi_{t+1}^{(\mathcal{S}+\mathcal{A})}\rangle =(U_N\otimes \openone_{\mathcal{A}})V_N |\Psi_{t}^{(\mathcal{S}+\mathcal{A})}\rangle,
\end{equation}
where $U_N$ is given by (\ref{multiU1}).

Let us assume for a moment that the state of $\mathcal{S}+\mathcal{A}$ is prepared in one of the basis states (we will consider superpositions in the following sections). Moreover, as before, let the string of k-qubits be initially prepared in the state $|00\ldots 0\rangle$. The future states of the string are fully determined by the particle number $N$ and the particle distribution at $t=0$. Note, that the state of the string at $t>0$ encodes the information about the system's past. For example, if after using $j<k$ qubits the string is in the state
\begin{equation}
|0 \ldots 11101\rangle,
\end{equation}
one can easily deduce that the last actions of the semipermeable barrier were: $\ldots$, {\it reflect}, {\it reflect}, {\it reflect}, {\it pass}, {\it reflect}. The situation when two particles meet at $x=x_0$ is not encoded in the string, since in this case the barrier takes no action. 


\subsection{Efficiency}

How many particles can be trapped using $k$ qubits? More precisely, how many particles can be trapped before the barrier starts to let the particles out of $\mathcal{R}$? The number of states in $\mathcal{R}$ is equal to $V_{\mathcal{R}}=2x_0 -1$ and the number of states in the remaining part of the box is equal to $\bar{V}_{{\mathcal{R}}}=V-V_{\mathcal{R}}$, where $V=2M$ is the total number of states. Let the initial number of particles in $\mathcal{R}$ be $N_0$ and the initial number of particles outside of $\mathcal{R}$ be $\bar{N}_0=N-N_0$, where $N$ is the total number of particles. Whenever a particle goes through the barrier and enters $\mathcal{R}$,  the density of particles inside $\mathcal{R}$ changes according to
\begin{equation}\label{Ln}
\rho(n) = \rho(n-1) + \frac{1}{V_{\mathcal{R}}} = \rho(0) + \frac{n}{V_{\mathcal{R}}} = \frac{N_0 + n}{V_{\mathcal{R}}} ,
\end{equation}
where $n$ counts the total number of particles that entered $\mathcal{R}$ from the outside. On the other hand, the density outside of $\mathcal{R}$ changes according to
\begin{equation}
\bar{\rho}(n) = \bar{\rho}(n-1) - \frac{1}{\bar{V}_{{\mathcal{R}}}} = \bar{\rho}(0) - \frac{n}{\bar{V}_{{\mathcal{R}}}} = \frac{\bar{N}_0 - n}{\bar{V}_{{\mathcal{R}}}}.
\end{equation}

Next, let us recall that a qubit from the string is used whenever: (i) a particle enters $\mathcal{R}$ from the outside, or (ii) a particle inside $\mathcal{R}$ reflects off the semipermeable barrier. Let $K(n)$ count the average number of reflections after the entry of the n-th particle and the entry of the (n+1)-th particle. The higher the density on one side, the higher the chance that the next particle will arrive at the barrier from this side. Therefore
\begin{equation}\label{Kn}
K(n) = \frac{\rho(n)}{\bar{\rho}(n)} + 1 = r\left(\frac{N_0+n}{\bar{N}_0 -n}\right) + 1,
\end{equation}
where the first term counts the average number of reflections and the last term corresponds to the entry of the (n+1)-th particle. In the above we introduced 
\begin{equation}
r \equiv \frac{\bar{V}_{\mathcal{R}}}{V_{\mathcal{R}}}.
\end{equation}

The total number of available qubits is $k$, therefore
\begin{equation}
k = \sum_{n=0}^{N_k-1} K(n),
\end{equation}
where $N_k$ is the total number of particles that entered through the barrier. We get
\begin{equation}\label{k}
k = N_k + r \sum_{n=0}^{N_k-1} \frac{N_0+n}{\bar{N}_0 -n}.
\end{equation}
It is useful to define $x\equiv \bar{N}_0 - n$, which leads to
\begin{eqnarray}
k &=& N_k + r \sum_{x=1}^{\bar{N}_0} \frac{N-x}{x} - r \sum_{x=1}^{\bar{N}_0-N_k} \frac{N-x}{x} \nonumber \\
&=& (1-r)N_k + r N \sum_{x=1}^{\bar{N}_0} \frac{1}{x} -  r N \sum_{x=1}^{\bar{N}_0-N_k} \frac{1}{x}. \label{sumk}
\end{eqnarray}
The sum can be approximated as
\begin{equation}
\sum_{x=1}^{\bar{N}_0} \frac{1}{x} \approx \log \bar{N}_0 + \gamma,
\end{equation}
where $\gamma \approx 0.57721$ is the Euler-Mascheroni constant. Therefore,
\begin{eqnarray}
k & \approx & (1-r) N_k + r N \log \left( \frac{\bar{N}_0}{\bar{N}_0 - N_k} \right)  \nonumber \\
& = & {\mathcal{O}}\left(N \log \left( \frac{\bar{N}_0}{\bar{N}_0 - N_k} \right) \right).
\end{eqnarray}
Note that if $N_k = \bar{N}_0$, the second sum in (\ref{sumk}) vanishes and 
\begin{equation}
k \approx (1-r) \bar{N}_0 + r N \gamma + r N \log  \bar{N}_0  = {\mathcal{O}}\left(N \log  \bar{N}_0 \right).
\end{equation}

Let us consider two particular limits. If the semipermeable barrier divides the box into two equal parts ($r=1$), the above formula simplifies to
\begin{equation}
k_{eq} \approx N \log \left( \frac{\bar{N}_0}{\bar{N}_0 - N_k} \right) .
\end{equation}   
On the other hand, if $\mathcal{R}$ is only a tiny fraction of the box (${V}_{\mathcal{R}} \ll \bar{V}_{\mathcal{R}}$ and ${N}_0 \ll \bar{N}_0 \approx N$), then we can make the following approximation. First, let us define 
\begin{equation}
R \equiv \frac{r}{N} = \frac{\bar{V}_{\mathcal{R}}}{V_{\mathcal{R}}{N}}.
\end{equation}
Then, let $N_k \ll  N$. In this case $\bar{N}_0 - N_k \approx \bar{N}_0 \approx N$ and the equation (\ref{k}) can be approximated as
\begin{eqnarray}
k_{\ll} &\approx & N_k + R \sum_{n=0}^{N_k-1} (N_0+n) \\
&=& (1+RN_0)N_k+R\frac{N_k(N_k -1)}{2} = {\mathcal{O}(N_k^2)}. \nonumber 
\end{eqnarray}


\subsection{Superpositions within ${\mathcal S}$}

Up to now we have considered the evolution in the computational basis. In this subsection we examine what happens if the initial state of $\mathcal{S}$ is a superposition of basis states. For the sake of this and the next subsections we relabel the basis states and assume that $k$ is large, i.e., the barrier does not run out of qubits throughout the evolution.

\begin{figure*}
  \includegraphics[trim=30 290 100 150,clip,scale=0.7]{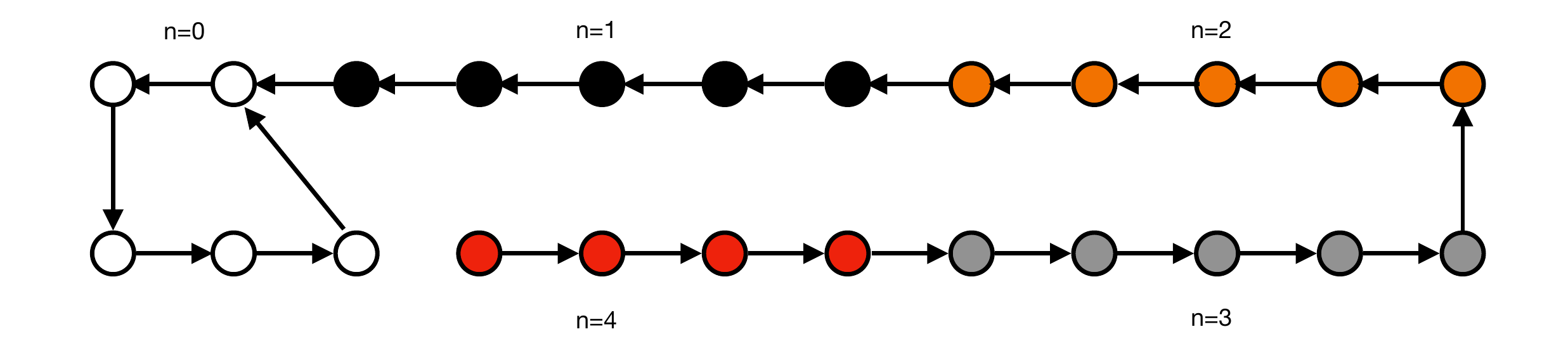}
  \caption{The division of basis states into sectors. Each but the last sector consists of $T$ states (see text).} \label{f3}
\end{figure*}

Note that a particle in any given state $|x,c \rangle$ is eventually going to bounce off the barrier after exactly $t$ steps ($1 \leq t \leq 2M$). Therefore, the basis states of $\mathcal{S}$ can be labeled $|t\rangle$ instead of $|x,c\rangle$, where $t$ denotes the number of evolution steps after which the particle bounces off the barrier. In particular, $|1\rangle \equiv |x_0,\rightarrow\rangle$, $|2\rangle \equiv  |x_0-1,\rightarrow\rangle$, and so on. The evolution transforms
\begin{eqnarray}
& & |t\rangle \rightarrow |t-1\rangle, ~~~~~~~~ (1 < t \leq 2M) \\
& & |1\rangle \rightarrow |T\rangle,
\end{eqnarray}
where $T=2x_0 -1$ and $|T\rangle \equiv |x_0 -1,\leftarrow\rangle$. The dynamics within $\mathcal{S}$ gradually reduces the number of states to $\{|t\rangle \}_{t=1}^{T}$. After the contraction the particle is bound to $\ {R}$ and its evolution is periodic. Thus, it will be convinient to write $|t \rangle$ as $|t+nT \rangle$ with $1 \leq t \leq T$ and $0 \leq n \leq K$, where $K = \lceil  2M/T \rceil $. The label $n$ divides all basic states of $\mathcal{S}$ into sectors in such a way that all states in a given sectors will be bound to $\mathcal{R}$ after time $nT$. This division is presented in Fig. \ref{f3}.

The basis states of $\mathcal{A}$ can be also relabelled. They are strings of zeros and ones whose possible distributions depend on the number of particles in $\mathcal{S}$. We first consider only one particle in which case these basis states can be relabelled as
\begin{equation}
|0\ldots0 \underbrace{1\ldots1}_{j}\rangle \equiv |j\rangle.
\end{equation}

The initial single-particle state of $\mathcal{S}+\mathcal{A}$ can be written as
\begin{equation}
|\Psi_0^{(\mathcal{S}+\mathcal{A})}\rangle = |\Psi_0^{(\mathcal{S})}\rangle\otimes |0\rangle
\end{equation}
with
\begin{equation}
|\Psi_0^{(\mathcal{S})}\rangle = \sum_{n=0}^{K-1}\sum_{t=1}^{T-1} \alpha_{t+nT} |t+nT \rangle   = \sum_{n=0}^{K-1} |\varphi^{(n)}_0 \rangle .
\end{equation}
In the above
\begin{equation}
|\varphi^{(n)}_0 \rangle = \sum_{t=1}^T \alpha_{t+nT}|t+nT\rangle
\end{equation}
is a superposition of states for which the particle bounces off the barrier at times $nT< t \leq (n+1)T$. More precisely, for $n=0$ we have $T$ states that are already inside $\mathcal{R}$, for $n=1$ we have the first group of $T$ states outside of $\mathcal{R}$, for $n=2$ we have the second group of $T$ states outside of $\mathcal{R}$ and so on. Note, that in $|\varphi^{(K-1)}_0 \rangle$ there might be elements corresponding to non-existing states $(t+(K-1)T > 2M)$. We therefore assume that $\alpha_t = 0$ if $t>2M$ (see Fig. \ref{f3}). 

After $T$ steps the system's state becomes
\begin{equation}
|\Psi_T^{(\mathcal{S}+\mathcal{A})}\rangle =  |\varphi^{(0)}_T \rangle \otimes |1\rangle + \left( \sum_{n=1}^{K-1} |\varphi^{(n)}_T \rangle \right) \otimes |0\rangle.
\end{equation}
After $2T$ steps it becomes
\begin{equation}
|\Psi_{2T}^{(\mathcal{S}+\mathcal{A})}\rangle =  |\varphi^{(0)}_{2T} \rangle \otimes |2\rangle +  |\varphi^{(1)}_{2T} \rangle \otimes |1\rangle + \left( \sum_{n=2}^{K-1} |\varphi^{(n)}_{2T} \rangle \right) \otimes |0\rangle
\end{equation}
and so on. In the above
\begin{equation}
|\varphi^{(n)}_{mT} \rangle = \sum_{t=1}^T \alpha_{t+nT}|t+\gamma_{n,m} T\rangle,
\end{equation}
where $\gamma_{n,m} = \max(n-m,0)$. Finally, after $KT$ steps the system's state is
\begin{equation}
|\Psi_{KT}^{(\mathcal{S}+\mathcal{A})}\rangle =   \sum_{n=0}^{K-1} |\varphi^{(n)}_{KT} \rangle \otimes |K-n\rangle,
\end{equation}
where
\begin{equation}
|\varphi^{(n)}_{KT} \rangle = \sum_{t=1}^T \alpha_{t+nT}|t\rangle.
\end{equation}
The reduced state of $\mathcal{S}$ is a mixture
\begin{equation}
\rho^{(\mathcal{S})}_{KT} = \sum_{n=0}^{K-1} |\varphi^{(n)}_{KT} \rangle \langle \varphi^{(n)}_{KT} |.
\end{equation}
As already mentioned, the subsequent evolution of $\rho^{(\mathcal{S})}_{KT} $ is periodic and $\rho^{(\mathcal{S})}_{mT} =\rho^{(\mathcal{S})}_{KT} $ for $m \geq K$.

It is clear that the action of the barrier may cause the state's coherence to drop. This is because the initial state is in general a superposition of $2M$ states, whereas the final state can be at most a superposition of $T$ states. To illustrate how coherence may be affected by the barrier let us consider two examples corresponding to two extreme cases. If for a given $n<K$
\begin{equation}
|\Psi_0^{(\mathcal{S})}\rangle=|\varphi^{(n)}_0 \rangle = \frac{1}{\sqrt{T}} \sum_{t=1}^T |t+nT\rangle,
\end{equation}
then $\rho^{(\mathcal{S})}_{KT}$ is a coherent pure state
\begin{equation}
 \rho^{(\mathcal{S})}_{KT} = |\varphi\rangle\langle \varphi|,
\end{equation}
where 
\begin{equation}
|\varphi \rangle = \frac{1}{\sqrt{T}} \sum_{t=1}^T |t \rangle.
\end{equation}
On the other hand, if for a given $t$
\begin{equation}
|\Psi_0^{(\mathcal{S})}\rangle=\sum_{n=0}^{K-1} |t+nT\rangle,
\end{equation}
then $\rho^{(\mathcal{S})}_{KT}$ is an even mixture (rank $K$) of mutually orthogonal states
\begin{equation}
 \rho^{(\mathcal{S})}_{KT} = \sum_{n=0}^{K-1} |t+nT\rangle\langle t+nT|.
\end{equation}


\subsection{Quantum correlations within ${\mathcal S}$}
 
The above change of coherence is also present when there is more than one particle in $\mathcal{S}$. We discuss in details the two-particle case. This time the basis states of $\mathcal{A}$ are of the form
\begin{equation}
|0\ldots0 \underbrace{1\ldots1}_{j} 0 \underbrace{1\ldots1}_{j'}\rangle \equiv |j,j'\rangle.
\end{equation}
where $j$ denotes the number of bounces off the barrier while there is only one particle inside $\mathcal{R}$, a single zero in between ones denotes the entry of the second particle into $\mathcal{R}$, and $j'$ denotes the number of bounces when both particles are inside $\mathcal{R}$. 

Let us consider the following initial two-particle state of $\mathcal{S}+\mathcal{A}$ 
\begin{eqnarray}
|\Psi_0^{(\mathcal{S}+\mathcal{A})}\rangle &=& \left(\sum_{1=t < t' }^{2M} \alpha_{t,t'} |t,t' \rangle \right) \otimes |0,0\rangle  \nonumber \\
&\approx& \left( \sum_{0=n\leq n'}^{K-1} |\varphi^{(n,n')}_0 \rangle \right) \otimes |0,0\rangle, \label{2part}
\end{eqnarray}
where $|t,t' \rangle \equiv a_{t'}^{\dagger}a_t^{\dagger} | \mathbf{0}\rangle$ corresponds to one particle bouncing off the barrier after $t$ steps and the other one after $t'$ steps. The sum over $t<t'$ takes into account the fact that the two particles are fermions and we do not distinguish which is which. As before, we assume that $\alpha_{t,t'} = 0$ for $t'>2M$. In addition,
\begin{equation}
|\varphi^{(n,n')}_0 \rangle = \sum_{t=1}^T\sum_{\tau=1}^{T-1} \beta_{t,\tau,n,n'}|t+nT,t+\tau+n'T\rangle,
\end{equation}
where $\beta_{t,\tau,n,n'} \equiv \alpha_{t+nT,t+\tau+n'T}$. Finally, in the above formula the sum over $\tau$ does not include the situation $\tau =T$, which is the reason why we used '$\approx$' symbol in (\ref{2part}). This is due to the fact that when $\tau =T$ the two fermions meet at the barrier, one bouncing off it and one entering $\mathcal{R}$. In this case only one fermion stays in $\mathcal{R}$. However, this can happen only for a small fraction of states, therefore, in order to keep things simple, we assume that $\alpha_{t+nT,t+n'T} = 0$ for all $n$ and $n'$. 

After $KT$ steps the two particles are trapped inside $\mathcal{R}$ and it is straightforward to show that the system is in the state
\begin{equation}
|\Psi_{KT}^{(\mathcal{S}+\mathcal{A})}\rangle =  \sum_{0=n\leq n'}^{K-1} |\varphi^{(n,n')}_{KT} \rangle  \otimes |n'-n,2(K-n')\rangle,
\end{equation}
where 
\begin{equation}
|\varphi^{(n,n')}_{KT} \rangle = \sum_{t=1}^T\sum_{\tau=1}^{T-1} \beta_{t,\tau,n,n'}|t,t+\tau \rangle,
\end{equation}
and the factor of two in $|n'-n,2(K-n')\rangle$ comes from the fact that for $t\geq n'T$ both particles bounce off the barrier. From now on the evolution within $\mathcal{S}$ is periodic and the reduced state is
\begin{equation}
\rho^{(\mathcal{S})}_{mT} =\rho^{(\mathcal{S})}_{KT} =  \sum_{0=n\leq n'}^{K-1} |\varphi^{(n,n')}_{KT} \rangle \langle \varphi^{(n,n')}_{KT} | 
\end{equation} 
for $m\geq K$.

The above dynamics may affect initial coherence and, as a result, quantum correlations between the particles. Since we assumed that particles are identical fermions, the notion of entanglement is no longer uniquely defined \cite{Corr1}. We adopt the formalism developed in \cite{Corr2,Corr3,Corr4,Corr5} and, instead of entanglement, speak of general quantum correlations. In particular, consider a two-fermion system described by
\begin{equation}
\sum_{1=i<j}^d \alpha_{ij} a_i^{\dagger}a_j^{\dagger} | \mathbf{0}\rangle,
\end{equation}
where each fermion can occupy one of $d$ states. Such a system is quantum correlated if there exists a basis 
\begin{equation}\label{basis}
\tilde{a}_k^{\dagger} = \sum_{i=1}^{d}\beta_{ik} a_i^{\dagger}
\end{equation}
in which its state can be represented as \cite{Corr2}
\begin{equation}\label{corr}
\sum_{j=1}^r \sqrt{\lambda_j} \tilde{a}^{\dagger}_{2j-1}\tilde{a}^{\dagger}_{2j}|\mathbf{0}\rangle.
\end{equation} 
for $r>1$ and $\sum_j \lambda_j = 1$. The number $r$ is known as {\it the Slater rank}. The basis states are divided into two separate sets, each occupied by exactly one fermion.  In the above formula the first set is labeled by odd numbers and the second one by even numbers, however any other labeling dividing the states into two separate groups would work. Therefore, the sets can be used to effectively distinguish the fermions and the insight from the standard entanglement theory can be used.  

Let us consider a particular example showing a clear decrease of quantum correlations. We assume $2M=KT$ and the initial state of the form 
\begin{equation}
|\Psi_0^{(\mathcal{S}+\mathcal{A})}\rangle = \left( \sum_{0=n}^{K-1} |\varphi^{(n,n)}_0 \rangle \right) \otimes |0,0\rangle, 
\end{equation}
for which 
\begin{equation}
|\varphi^{(n,n)}_0 \rangle = \frac{1}{\sqrt{K(T-1)}}\sum_{t=1}^{(T-1)/2} |2t-1+nT,2t+nT\rangle,
\end{equation}
This state is highly quantum correlated within $\mathcal{S}$ and the corresponding Slater rank is $r=K(T-1)/2$. However, after $KT$ steps the reduced state of $\mathcal{S}$ becomes
\begin{equation}
\rho^{(\mathcal{S})}_{KT} =  |\chi \rangle \langle \chi |,
\end{equation} 
where 
\begin{equation}
|\chi \rangle = \frac{1}{\sqrt{K(T-1)}}\sum_{t=1}^{(T-1)/2} |2t-1,2t \rangle.
\end{equation}
The Slater rank of this state is $r'=r/K=(T-1)/2$. 
 

\subsection{Superpositions within ${\mathcal A}$}
 
Finally, we present a particular effect that emerges when $\mathcal{A}$ is in a superposition of basis states. First we focus on a single scattering event at the barrier which involves a single particle in $\mathcal{S}$ and a single qubit in $\mathcal{A}$. The qubit is in a state $\alpha |0\rangle + \beta|1\rangle$ and we assume that at time $t$  a single particle arrives at the barrier from the right. At that moment the state of $\mathcal{S}+\mathcal{A}$ is
 \begin{equation}
 |\Psi_t^{(\mathcal{S}+\mathcal{A})}\rangle = |x_0,\leftarrow\rangle \otimes (\alpha |0\rangle + \beta|1\rangle).
 \end{equation}
 The evolution (recall Eqs. (\ref{W}) and (\ref{wallreversed})) transforms this state into 
  \begin{equation}
 |\Psi_{t+1}^{(\mathcal{S}+\mathcal{A})}\rangle = (\alpha |x_0-1,\leftarrow\rangle + \beta  |x_0+1,\rightarrow\rangle) \otimes  |0\rangle.
 \end{equation}
 The superposition within $\mathcal{A}$ is transferred to $\mathcal{S}$. The barrier acts as a programmable beam splitter that transmits with the probability amplitude $\alpha$ and reflects with the probability amplitude $\beta$. After this operation the qubit is transformed into the state $|0\rangle$, therefore its initial coherence can be considered as a resource that is consumed during the beam splitting. 

If the particle arrived from the left, the state
\begin{equation}
 |\Psi_t^{(\mathcal{S}+\mathcal{A})}\rangle = |x_0,\rightarrow\rangle \otimes (\alpha |0\rangle + \beta|1\rangle)
 \end{equation}
 would be transformed into
 \begin{equation}
 |\Psi_{t+1}^{(\mathcal{S}+\mathcal{A})}\rangle = (\alpha |x_0-1,\leftarrow\rangle + \beta  |x_0+1,\rightarrow\rangle) \otimes  |1\rangle.
 \end{equation}
 This shows that the barrier does not act as an ordinary beam splitter. The particle's output state does not depend on its input state. Instead, the qubit's output state depends on the particle's input state. In other words, the particle's past is encoded on the qubit.    
 
This effect can be easily extended to more qubits and more particles. For example, an entangled N-qubit state $\alpha|0\rangle^{\otimes N}+ \beta|1\rangle^{\otimes N}$ allows for a collective beam splitting of N particles, i.e., creating a superposition of all particles going right and all particles going left. In such a process the N-partite entanglement of qubits is transferred into N-partite entanglement of particles.


\section{Summary and discussion} 

The answers to the four questions raised at the beginning are the following

\begin{enumerate}

\item The unitary dynamics capable of trapping particles in a finite region $\mathcal{R}$ consists of a particles' free evolution and an interaction between the particles and an auxiliary system $\mathcal{A}$. The interaction prevents the particles from leaving $\mathcal{R}$. In addition, the state of $\mathcal{A}$ changes each time a particle enters $\mathcal{R}$ or tries to leave it.


\item The trapping time and the number of particles that can be trapped inside $\mathcal{R}$ depend on the size of $\mathcal{A}$. If $\mathcal{A}$ consists of $k$ qubits, the semipermeable barrier stops functioning properly after $k$ particle entries or exit attempts. On the other hand, the number of particles $N_k$ that can be trapped before the barrier stops to work properly depends on the total number of particles $N$ and the number of particles $\bar{N}_0$ that are initially outside of $\mathcal{R}$.  The relation between these numbers scales as $k= {\mathcal{O}}\left(N \log \left( \frac{\bar{N}_0}{\bar{N}_0 - N_k} \right) \right)$.

\item Superpositions and quantum correlations of the particles are affected by the action of the semipermeable barrier. This is due to an entangling property of the interaction between the particles and $\mathcal{A}$. In general, any initial pure state of the particles that extends over distances larger than the number of states in $\mathcal{R}$ looses some of its coherence and some of its quantum correlations due to the trapping. 

\item If the initial state of $\mathcal{A}$ is in a superposition of basis states, the semipermeable barrier acts as a programmable beam splitter that transfers coherence from $\mathcal{A}$ to the particles. If the initial state of  $\mathcal{A}$ is entangled, the barrier causes a collective beam splitting and the entanglement gets transferred to the particles.

\end{enumerate}

Although we considered a particular discrete space-time model, our results should apply to arbitrary reversible evolutions of bipartite systems $\mathcal{S}+\mathcal{A}$ in which the state space of $\mathcal{S}$ gets contracted. For example, it is natural to expect that in all finite physical systems, governed by an autonomous evolution, any observable contraction of a subsystem's state space is only temporary. 


The studies in this work highlight a peculiar property of the quantum Maxwell's demon. The wave-particle duality allows for an alternative view of the demon's workings. Instead of trapping particles, the demon has to trap a wave inside $\mathcal{R}$. This wave is described by both, amplitudes and phases. The phases give rise to coherences between different positions. The action of the demon is twofold. On one hand, it gathers all amplitudes inside $\mathcal{R}$. On the other, it erases some coherences. It seems that the partial erasure of coherences is the price one has to pay for the wave's localisation.

Finally, the programmable beam splitting property of our model may find an application in studies on composite particles \cite{Comp1,Comp2,Comp3}. More precisely, entanglement in $\mathcal{A}$ allows for a collective beam splitting of a number of particles, which may be particularly useful to observe interference of composite objects \cite{Comp4}.   


\section{Acknowledgements} 

This research is supported by the Polish National Science Centre (NCN) under the Maestro Grant no. DEC-2019/34/A/ST2/00081. 



\subsection{Appendix A}

Let us consider a simple example proving that a semipermeable barrier generates an irreversible evolution. Assume that $x_0 = 2$ and that the semipermeable barrier lets the particles coming in from the right. In addition, let $M\geq 6$. First, let us prepare a particle in the state $|\psi^{(a)}_0\rangle = |3,\leftarrow\rangle$.  After one step the particle enters $\mathcal{R}$ and is in the state $|\psi^{(a)}_{1}\rangle = |2,\leftarrow\rangle$. After two steps it is in the state $|\psi^{(a)}_{2}\rangle = |1,\leftarrow\rangle$. After three steps it is in the state $|\psi^{(a)}_{3}\rangle = |1,\rightarrow\rangle$. Finally, after four steps the particle bounces off the semipermeable barrier at $x_0$ and is in the state $|\psi^{(a)}_{4}\rangle = |2,\leftarrow\rangle$. In general, the trapped particle follows a periodic evolution whose period is $2x_0 -1$. Note that the state $|x_0,\rightarrow\rangle$ is immediately transformed by the semipermeable barrier into $|x_0,\leftarrow\rangle$ (see Appendix B).

Next, let us prepare a particle in the state $|\psi^{(b)}_{0}\rangle = |6,\leftarrow\rangle$. After four steps the particle enters $\mathcal{R}$ and is in the state $|\psi^{(b)}_{4}\rangle = |2,\leftarrow\rangle$. The second preparation is perfectly distinguishable from the first one, since $\langle \psi^{(a)}_0|\psi^{(b)}_0 \rangle=0$. If the above evolution were reversible, the following would hold $\langle \psi^{(a)}_t|\psi^{(b)}_t \rangle=0$ for all $t$. However, $\langle \psi^{(a)}_{4}|\psi^{(b)}_{4} \rangle=1$, therefore the evolution is irreversible. 


\subsection{Appendix B}

The action of a semipermeable barrier can be described by Kraus operators. For example, a particle at $x_0$ can evolve according to:
\begin{equation}
\rho_{t+1} = U K_1 \rho_t K_1^{\dagger} U^{\dagger} + U K_2 \rho_t K_2^{\dagger} U^{\dagger}
\end{equation}  
where $U$ is given by (\ref{U}), $\rho_t$ is the particle's density matrix at time $t$ and the two Kraus operators are 
\begin{eqnarray}
K_1 &=& (\openone_x - |x_0\rangle\langle x_0|)\otimes \openone_c + |x_0,\leftarrow\rangle\langle x_0,\rightarrow|,  \\
K_2 &=& |x_0,\leftarrow\rangle\langle x_0,\leftarrow|.
\end{eqnarray}
In the above $\openone_x$ and $\openone_c$ are the identity operators on the position and coin subspaces, respectively.

The above operators do not exactly correspond to the reduced dynamics generated by the unitary operator (\ref{V}) or (\ref{VN}). The only difference is that the unitary operators use a new ancillary qubit only when a particle is reflected or transmitted through a barrier. On the other hand, the Kraus operators stem from a unitary operator that uses a new ancillary  qubit at each time step. More precisely, the above Kraus operators stem from the following version of the operator (\ref{V})
\begin{eqnarray}
V &=& (\openone_{\mathcal{S}} \otimes T)(W\otimes\openone^{\otimes k-1}), \nonumber
\end{eqnarray}
where $T$ is defined in (\ref{T}) and $W$ is defined in (\ref{W}) and (\ref{wallreversed}).


\subsection{Appendix C}

Let us apply the above Kraus operators to two identical fermions. We use the first quantization picture. Let us assume $x_0 = 2$, $M \geq 7$ and the two fermions in the initial state $\rho_0 = |\psi_0\rangle\langle \psi_0|$, where
\begin{equation}
|\psi_0\rangle = \frac{1}{\sqrt{2}}(|3,\leftarrow\rangle\otimes|7,\leftarrow\rangle -|7,\leftarrow\rangle\otimes|3,\leftarrow\rangle).
\end{equation}
The evolution is given by
\begin{equation}
\rho_{t+1} = U\left(\sum_{i=1}^3 \kappa_i \rho_t \kappa_i^{\dagger} \right) U^{\dagger},
\end{equation}
where, due to indistinguishability, $\kappa_i$ are symmetrized Kraus operators
\begin{eqnarray}
\kappa_1 &=& K_1\otimes K_1, \nonumber \\
\kappa_2 &=& K_2\otimes K_2, \nonumber \\
\kappa_3 &=& K_1\otimes K_2 + K_2\otimes K_1.
\end{eqnarray}
Note that $K_2^{\dagger}K_1 = K_1$ and $K_1^{\dagger}K_2 = 0$, therefore
\begin{equation}
\sum_{i=1}^{3}\kappa_i^{\dagger}\kappa_i =  \sum_{k,l=1}^2 K_k^{\dagger}K_k \otimes K_l^{\dagger}K_l = \openone\otimes\openone.
\end{equation}
After one step the system is in the state $\rho_1=|\psi_1\rangle\langle\psi_1|$
\begin{equation}
|\psi_1\rangle = \frac{1}{\sqrt{2}}(|2,\leftarrow\rangle\otimes|6,\leftarrow\rangle -|6,\leftarrow\rangle\otimes|2,\leftarrow\rangle),
\end{equation}
and after six steps the system is in the state $\rho_{6} = |\psi_{6}\rangle\langle\psi_{6}|$
\begin{equation}
|\psi_{6}\rangle = \frac{1}{\sqrt{2}}(|1,\rightarrow\rangle\otimes|1,\leftarrow\rangle -|1,\leftarrow\rangle\otimes|1,\rightarrow\rangle).
\end{equation}
This proves that particles remain antisymmetrized both, inside and outside $\mathcal{R}$.


\end{document}